\documentclass[aps,a4paper,twocolumn]{revtex4-2}
\usepackage{amsmath}
\usepackage{graphicx}
\usepackage{subfigure}
\usepackage[usenames,dvipsnames]{color}
\definecolor{darkblue}{RGB}{0,0,196}
\usepackage[colorlinks=true,linkcolor=darkblue,citecolor=darkblue,urlcolor=darkblue]{hyperref}
\usepackage{setspace}
\usepackage{hyperref}
\usepackage{xcolor}
\hypersetup{
  colorlinks   = true, 
  urlcolor     = red, 
  linkcolor    = blue, 
  citecolor   = blue 
}
\usepackage{graphicx}
\usepackage{amsmath,bbm}
\usepackage{amssymb,bm}
\usepackage{yfonts}
\usepackage{comment}

\usepackage[usenames,dvipsnames]{color}
\definecolor{darkblue}{RGB}{0,0,196}
\usepackage[colorlinks=true,linkcolor=darkblue,citecolor=darkblue,urlcolor=darkblue]{hyperref}

\begin{document}
\title{Partial pressure and susceptibilities of charmed sector in the van der Waals hadron resonance gas model}

\author{Kangkan Goswami$^1$}
\author{Kshitish Kumar Pradhan$^1$}
\author{Dushmanta Sahu$^2$}
\author{Raghunath Sahoo$^1$}
\email{Corresponding Author: Raghunath.Sahoo@cern.ch}

\affiliation{$^1$Department of Physics, Indian Institute of Technology Indore, Simrol, Indore 453552, India}
\affiliation{$^2$Instituto de Ciencias Nucleares, UNAM, Apartado Postal 70-543, Coyoacán, 04510, México City, México}


\begin{abstract}

We investigate the general susceptibilities in the charm sector by using the van der Waals hadron resonance gas model (VDWHRG).We argue that the ideal hadron resonance gas (HRG), which assumes no interactions between hadrons, and the excluded volume hadron resonance gas (EVHRG), which includes only repulsive interactions, fail to explain the lQCD data at very high temperatures. In contrast, the VDWHRG model, incorporating both attractive and repulsive interactions, extends the degree of agreement with lQCD up to nearly 180 MeV. We estimate the partial pressure in the charm sector and study charm susceptibility ratios in a baryon-rich environment, which is tricky for lattice quantum chromodynamics (lQCD) due to the fermion sign problem. Our study further solidifies the notion that the hadrons shouldn't be treated as non-interacting particles, especially when studying higher order fluctuations, but rather one should consider both attractive and repulsive interactions between the hadrons. 
 
\end{abstract}
\maketitle
\section{Introduction}

Understanding the behavior of matter formed in ultra-relativistic collisions is one of the most interesting areas of research currently. While colliders such as the Large Hadron Collider (LHC) at CERN and the Relativistic Heavy Ion Collider (RHIC) at BNL give us important information through experimental data, on the other hand, theoretical and phenomenological models clarify our views of the medium in the areas where colliders have limitations. Lattice Quantum Chromodynamics (lQCD) is a high-temperature first-principles QCD theory, which is one of the most important tools to study the QCD phase diagram. It also estimates the thermodynamic and transport properties of the matter at high temperature ($T$) and vanishing baryochemical potential ($\mu_{\rm B}$). The lQCD predicts that the degrees of freedom of the system change smoothly from hadronic to partonic degrees of freedom, approaching the Stefan-Boltzmann limit at zero baryo-chemical potential, quantified by the study of scaled energy density ($\varepsilon/T^{4}$)~\cite{Borsanyi:2010cj, Borsanyi:2013bia}. However, the applicability of lQCD breaks down at high $\mu_{\text{B}}$ due to the fermion sign problem~\cite{HotQCD:2014kol}.  There are a lot of alternatives to lQCD theory in the deconfined medium sector, such as the Nambu-Jona-Lasinio model~\cite{Marty:2013ita, Ghosh:2019ubc, Dwibedi:2025boz}, the Polyakov loop Nambu-Jona-Lasino (PNJL) model~\cite{Ratti:2005jh, Ghosh:2014vja, Ferreira:2014kpa, Goswami:2023eol, Costa:2008dp}, the Dynamical quasi-particle model (DQPM)~\cite{Soloveva:2019xph, Cassing:2007nb, Berrehrah:2015vhe}, functional renormalization group theory (FRG)~\cite{Wetterich:1992yh, Dupuis:2020fhh}, Color String Percolation model (CSPM)~\cite{Sahu:2020mzo, Sahu:2020nbu, Goswami:2022szb}, etc, which are used to estimate various thermodynamic and transport properties. On the other hand, in the low temperature regime, where hadronic degrees of freedom dominate, the hadron resonance gas (HRG) model is a robust tool. Despite being very simplistic in nature, it can explain the lQCD thermodynamic estimations up to $T \simeq 150$ MeV~\cite{Karsch:2003vd}. The HRG model also succeeds in explaining the hadronic yields from the experimental data~\cite{Andronic:2005yp}. However, it lacks in explaining the higher-order conserved charge fluctuations estimations by lQCD~\cite{Borsanyi:2011sw, HotQCD:2017qwq}. There have been improvements to the HRG model, more notably the excluded volume hadron resonance gas (EVHRG) model, which assumes the hadrons to have finite radii mimicking a negative pressure in the system~\cite{Rischke:1991ke, Yen:1997rv}. EVHRG explains the thermodynamic and higher-order fluctuations in lQCD data better than HRG, although some disagreements can be seen at the higher temperature regime. Recently, a new approach was developed where the authors assumed van der Waals-like interactions between the hadrons in the medium~\cite{Vovchenko:2016rkn}. This model takes both attractive and repulsive interactions among hadrons, whose interplay helps to better explain the lQCD data, even up to 180 MeV. This model can also explain the conserved charge fluctuations to a good extent. A lot of studies have been done exploring the validity and further predictions from the VDWHRG model, explaining thermodynamics and transport properties of the hadronic matter~\cite{Pradhan:2022gbm, Pradhan:2023rvf, Sahoo:2023vkw, Pradhan:2023etz, Singh:2025rwc}.

A unique way to study the medium properties is through the study of heavy-flavor hadrons. Due to their higher masses, the charm and bottom quarks are produced relatively early in the system from the hard scatterings. They witness the whole medium evolution from the quark-gluon plasma to the hadronization phase, where they combine with other quarks to produce hidden and open charm or bottom hadrons. It is possible to gain valuable insights into the produced medium by studying the yields of such heavy-flavor hadrons. A lot of studies have been done to understand the production, polarization, and even effects like collective flow in the heavy-flavor sector. On the other hand, charm fluctuations, quantified through susceptibilities, are particularly sensitive to the degrees of freedom in the medium. These susceptibilities are derived from the derivatives of the pressure of the system with respect to the charm chemical potential, and they encode information about the correlations and higher-order moments of charm quantum number distributions. In the hadronic phase, the charm sector is dominated by charmed mesons (e.g., $D$ mesons) and baryons (e.g., $\Lambda_{c}$ baryons), while in the deconfined phase, charm quarks and their interactions with the medium become the primary contributors. The transition between these regimes is expected to show distinct signatures in the behavior of charm fluctuations, making them promising observables for identifying the crossover or critical point in the QCD phase diagram. Recently, a few studies explored the behavior of the charmed degrees of freedom by estimating the charm fluctuations using lQCD calculations at zero baryochemical potential~\cite{Bazavov:2023xzm, Kaczmarek:2025dqt, Sharma:2024edf}. They also compare how the HRG model behaves with respect to lQCD in the charm sector. A visible disagreement can be observed between the two models, even after considering the undiscovered charm states through the Quark model~\cite{Bazavov:2014yba}. These discrepancies suggest that interactions among hadrons, as well as the possible emergence of deconfined charm degrees of freedom, play significant roles at temperatures beyond 150 MeV.

In Ref.~\cite{Goswami:2023hdl}, we have reported how interaction in the hadronic medium, specifically the VDW interactions, along with accounting for the undiscovered hadronic states, improve the agreement between the hadronic and lQCD theory. We have also explored how the charm quantum number affects the diffusion matrix by using the VDWHRG model~\cite{Goswami:2024hfg}, where we predict that charm fluctuation can be a better probe to study QCD phase transition and critical point due to the low diffusion of the charm quantum number. In the current article, we extend our previous studies and explain the estimation of partial pressure and higher-order charm susceptibilities by using the VDWHRG model. In addition to this, we also explore these quantities in high baryon-rich environments. The behavior of charm fluctuations at high baryon densities, which is relevant for the beam energy scan programs at RHIC and future facilities like FAIR and NICA, remains an open question due to the limitations of lQCD at finite $\mu_{B}$. Phenomenological models like VDWHRG, supplemented with additional constraints from experimental data, offer an interesting pathway to explore this uncharted territory. Additionally, this article is organized as follows. In the formulation section~\ref{Formulation}, we give a brief introduction to HRG, EVHRG, and VDWHRG models. In section \ref{Results}, we discuss the results. Finally, in section~\ref{sum}, we summarize our findings.

\section{Formulation}
\label{Formulation}
\subsection{Ideal Hadron Resonance Gas Model}

The HRG model is a theoretical framework used to describe the thermodynamic properties of strongly interacting matter in the hadronic phase. The ideal HRG (IHRG) model assumes that the hadronic matter is in thermal equilibrium, and all the hadronic states are point-like particles with no interactions between them. The grand canonical partition function in the IHRG model is defined as,

\begin{equation}
{\rm ln}Z_{i}^{id} = \pm\frac{Vg_{i}}{2\pi^{2}} \int_{0}^{\infty} p_{i}^{2}dp_{i} ~{\rm ln}(1\pm exp[-(E_{i} - \mu_{i})/T])
\end{equation}
 where, $g_{i}$, $p_{i}$, $m_{i}$ and $E_{i} = \sqrt{p_{i}^{2} + m_{i}^{2}}$ are the degeneracy, momentum, mass and energy of the $i$th hadron respectively. The $\pm$ signs correspond to fermions and bosons in the system, and $\mu_{i}$ is the total chemical potential of the system given as,
\begin{equation}
    \mu_{i} = B\mu_{\rm{B}} + S\mu_{\rm{S}} + Q\mu{\rm{Q}} + C\mu_{\rm{C}},
\end{equation}
where B, S, Q, and C are baryon, strangeness, electric charge, and charm quantum numbers, respectively. Finally, the pressure $P_{i}$ in IHRG is given as,

 \begin{equation}
     P_{i}^{id} = \pm \frac{Tg_{i}}{2\pi^{2}} \int_{0}^{\infty} p_{i}^{2}dp_{i} ~{\rm ln}(1\pm exp[-(E_{i} - \mu_{i})/T])
 \end{equation}

\subsection{Excluded Volume Hadron Resonance Gas model}
The EVHRG model is an extension of the IHRG model, which takes into account the finite size of hadrons. This finite volume of hadrons exerts an outward pressure in the system, which behaves as a repulsive interaction between the hadrons. In a thermodynamically consistent EVHRG model, the pressure can be written as,
\begin{equation}
    P^{EV}(T,\mu) = P^{id}(T,\mu^{*})
\end{equation}
Here, $\mu^{*}$ is the effective chemical potential given as,

\begin{equation}
    \mu^{*} = \mu - bP^{EV}(T, \mu).
\end{equation}
We take the repulsive parameter, $b = \frac{16}{3}\pi r^{3}$, where $r$ is the hardcore radius taken as, $r_{M} = 0.2~\text{fm}$ for mesons and $r_{B(\Bar{B})}=0.62~\text{fm}$ for baryons~\cite{Sarkar:2018mbk}.

\subsection{van der Waals Hadron Resonance Gas model}
The VDWHRG model takes both attractive and repulsive interactions into consideration. In the GCE, the van der Waals modified pressure is given as~\cite{Vovchenko:2015vxa, Vovchenko:2015pya},
\begin{equation}
    P(T,\mu) = P^{id}(T, \mu^{*}) - an^{2}(T,\mu).
    \label{eq_prs_vdw}
\end{equation}
Here, $n$ denotes the number density obtained from the VDWHRG model, and $\mu^{*}$ represents the effective chemical potential can expressed as,
\begin{equation}
    n(T,\mu) = \frac{\sum_{i}n_{i}^{id}(T,\mu^{*})}{1+b\sum_{i}n_{i}^{id}(T,\mu^{*})},
    \label{eq_nvdw}
\end{equation}
\begin{equation}
        \mu^{*} = \mu - bP(T,\mu) - abn^{2}(T,\mu) + 2an(T,\mu).
\end{equation}
For VDWHRG, the attractive interactions are encoded through the parameter $a$. For this work, we take $a=0.926~\rm{GeV~fm^{3}}$~\cite{Sarkar:2018mbk}. The repulsive interactions are taken care of by introducing the excluded volume effect with the parameter $b$, as defined previously.

\subsection{Partial pressure and Charm susceptibilities}

We calculate the generalized charm susceptibilities, which can be estimated from the derivative of medium pressure with respect to the chemical potentials, given as
\begin{equation}
\label{equation_susceptibility}
\chi^{BQSC}_{klmn} = \frac{\partial^{k+l+m+n}[P(\mu_{B}, \mu_{Q}, \mu_{S}, \mu_{C})/T^{4}]}{\partial \hat{\mu}_{B}^{k} ~\partial \hat{\mu}_{Q}^{l} ~\partial \hat{\mu}_{S}^{m} ~\partial \hat{\mu}_{C}^{n}} \bigg |_{\mu = 0},
\end{equation}
where $k$, $l$, $m$, and $n$ are integers denoting the order of the derivative. For a system of hadrons, the total pressure can be understood as the sum of the partial pressure of the charm sector and the partial pressure of the hadrons with $|C|=0$. Moreover, the partial charm pressure can be expressed in terms of the partial pressure of charmed baryons and mesons~\cite{Bazavov:2023xzm, Sharma:2024edf, Kaczmarek:2025dqt},
\begin{equation}
    P^{C}(T, \mu) = P_{B}^{C}(T, \mu) + P_{M}^{C}(T, \mu).
    \label{equation_charm_partial_pressure}
\end{equation}
Furthermore, the partial pressure of the charmed baryons and charmed mesons can be expressed as a linear combination of the charm susceptibilities~\cite{Bazavov:2023xzm, Sharma:2024edf, Kaczmarek:2025dqt}.
\begin{subequations}\label{equation_partial_pressure}
\begin{align}
	P_B^{C} &= \frac{1}{2} \left( 3\chi_{22}^{BC} - \chi_{13}^{BC} \right) \, , \label{eq:PC_B} \\
	P_M^{C} &= \chi_4^{C} + 3\chi_{22}^{BC} - 4\chi_{13}^{BC} \, . \label{eq:PC_M}
\end{align}
\end{subequations}

Here, $P_{B}^{C}$ and $P_{M}^{C}$ are the partial charm baryon and meson pressure, respectively. The charm susceptibilities $\chi_{4}^{C}$, $\chi_{22}^{BC}$, and $\chi_{13}^{BC}$ are estimated using Eq.~\ref{equation_susceptibility}.
Similarly, for the charmed-strange sector, we can express the partial pressure as~\cite{Kaczmarek:2025dqt},
\begin{subequations}\label{equation_partial_strange}
\begin{align}
    P_M^{C, S=1} &= \chi_{13}^{SC} - \chi_{112}^{BSC} \, , \label{eq:partial_strange_a} \\
    P_B^{C, S=1} &= \chi_{13}^{SC} - \chi_{22}^{SC} - 3\chi_{112}^{BSC} \, , \label{eq:partial_strange_b} \\
    P_B^{C, S=2} &= \frac{1}{2}\!\left( 2\chi_{112}^{BSC} + \chi_{22}^{SC} - \chi_{13}^{SC} \right) \, . \label{eq:partial_strange_c}
\end{align}
\end{subequations}
For the charged charm sector, the partial pressure for hadrons carrying the electric charge, $|Q|=2$, we write the partial pressure as~\cite{Sharma:2024edf}
\begin{equation}
    P_C^{Q=2} = \frac{1}{8} \left( 2\chi_{13}^{QC} - 5\chi_{22}^{QC} + 3\chi_{31}^{QC} \right).
\label{equation_partial_charged}
\end{equation}

The partial pressures of the charm sector, as defined in Eqs.~\ref{equation_charm_partial_pressure}-\ref{equation_partial_charged}, are fundamentally connected to lattice QCD (lQCD) calculations through a Taylor expansion of the QCD pressure around vanishing chemical potentials. In lQCD, the generalized susceptibilities $\chi^{BQSC}_{klmn}$ are computed numerically as derivatives of the partition function at $\mu = 0$, serving as coefficients in this expansion. These susceptibilities quantify fluctuations and correlations among quantum numbers (baryon number $B$, charge $Q$, strangeness $S$, and charm $C$), with diagonal terms like $\chi^C_4$ measuring pure charm fluctuations and off-diagonal terms like $\chi^{BC}_{22}$ capturing baryon-charm correlations. The linear combinations of susceptibilities in Eqs.~\ref{equation_partial_pressure}-\ref{equation_partial_charged} effectively project out contributions from specific hadron classes (e.g., charmed baryons vs. mesons), enabling direct comparison between hadronic models (HRG, VDWHRG) and lQCD results. This approach is particularly powerful because it avoids the sign problem in lQCD at finite density while providing a clear probe of how charm hadronization depends on the medium properties.

\begin{figure*}
    \centering
    \includegraphics[width=0.45\linewidth]{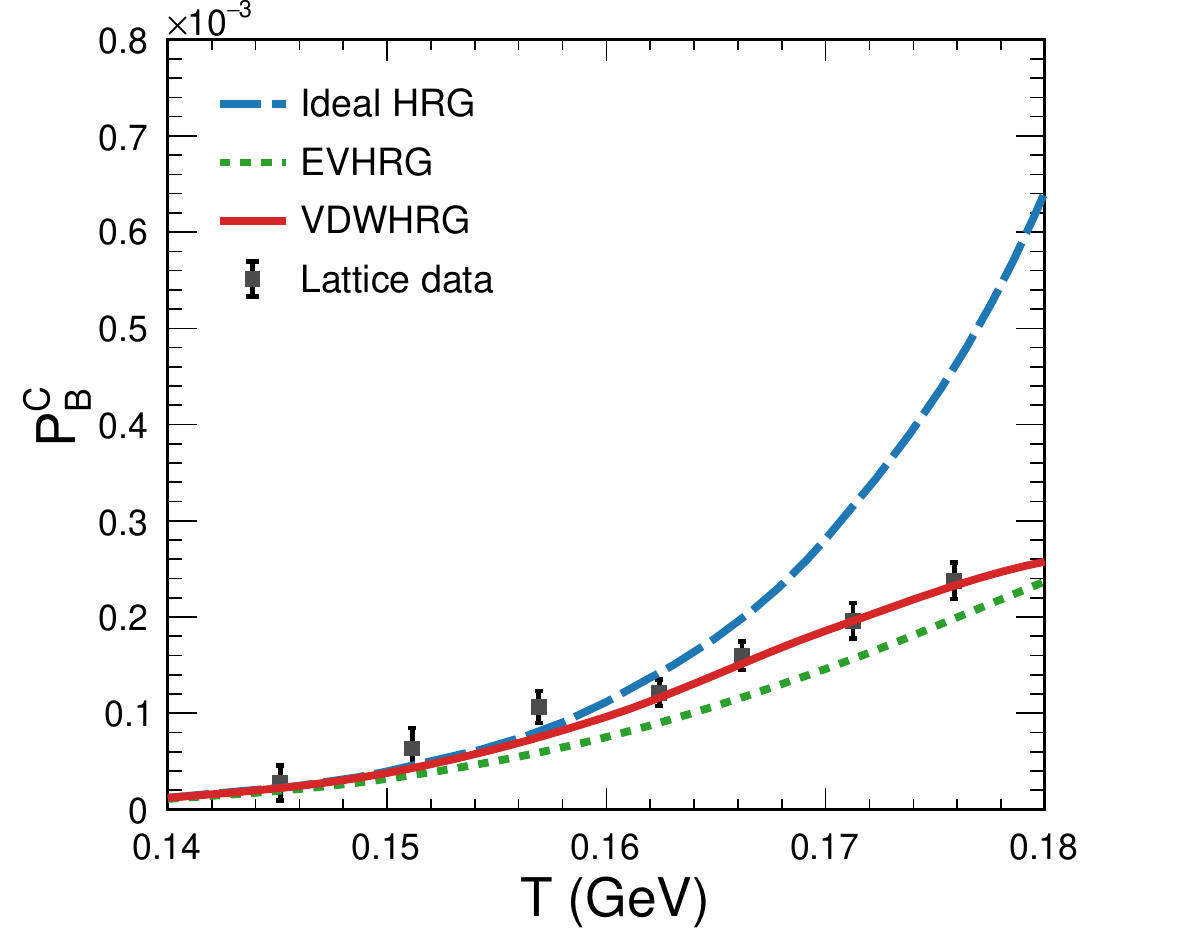}
    \includegraphics[width=0.45\linewidth]{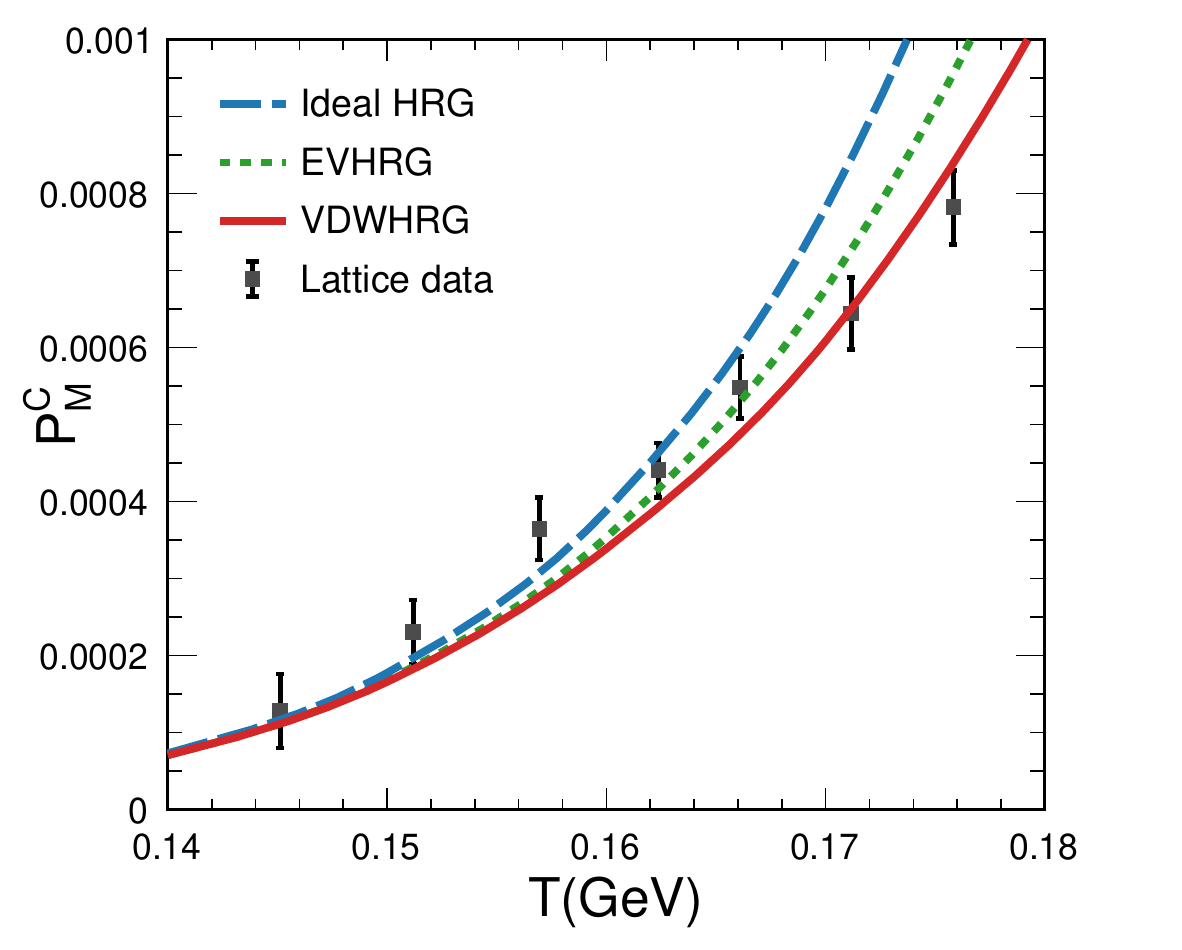}
    \caption{Partial pressure of charmed baryons (left panel) and charmed mesons (right panel) as functions of temperature at vanishing baryo-chemical potential. The markers are for lattice QCD results~\cite{Kaczmarek:2025dqt}, the solid red line is for VDWHRG, the dashed blue line is for ideal HRG, and the dotted green line is for EVHRG.}
    \label{fig1}
\end{figure*}

Moreover, the parametrization given in Ref.\cite{Cleymans:2005xv} allows us to study the charm susceptibilities as a function of center-of-mass energy. The well-established parametrization is given as, 
\begin{equation}
    T(\mu_{B}) = q_{1} - q_{2}\mu_{B}^{2} - q_{3}\mu_{B}^{4},
\end{equation}
\begin{equation}
    \mu_{B}(\sqrt{s_{\text{NN}}}) = \frac{q_{4}}{1+q_{5}\sqrt{s_{\text{NN}}}},
\end{equation}
where, $q_{1} = 0.166~\text{GeV}$, $q_{2} = 0.139~\text{GeV}^{-1}$, $q_{3} = 0.053~\text{GeV}^{-3}$, $q_{4} = 1.308~\text{GeV}$, and $q_{5} = 0.273~\text{GeV}^{-1}$. These parameters are obtained using freeze-out criteria based on the ideal HRG model. Although similar estimations have been made using the EVHRG and VDWHRG model, the resulting parameter variations are negligible~\cite{Poberezhnyuk:2019pxs, Behera:2022nfn, Tiwari:2011km}. Moreover, one should take note of the fact that this freeze-out parametrization determined from the yields of light-flavor hadrons. We work with the assumption that the charm hadrons decouple from the medium at the same temperature as light hadrons. For a comparison on equal footing, we employ the experimental $p_{\rm{T}}$ and $\eta$ cuts in the estimation of charm susceptibilities. The pressure integral in the hadronic models modifies as~\cite{Alba:2014eba},
\begin{equation}
\begin{aligned}
P_{i}^{id} = \pm \frac{T g_{i}}{4\pi^{2}}
&\int_{-\eta^{min}}^{\eta^{max}} d\eta 
\int_{p_{\rm T}^{min}}^{p_{\rm T}^{max}} dp_{\rm T} ~ p_{\rm T,i}^{2} \cosh{\eta}  \\
&\times {\rm ln}\!\left(1 \pm e^{-(E^{i}(p_{\rm{T}}, \eta)-\mu_{i})/T}\right),
\end{aligned}
\end{equation}
here, $E^{i}(p_{\rm{T}}, \eta)$ is the energy of the $i^{th}$ hadron given as, $E_{i}(p_T,\eta) = \sqrt{p_{\rm T,i}^{2}\cosh^{2}\eta + m_{i}^{2}}
$. Moreover, the chemical potentials are constrained by imposing $n_{Q}/n_{B} = 0.4$, where $n_{Q}$ and $n_{B}$ are the net electric charge and net baryon densities, along with strangeness and charm neutrality. Furthermore, we incorporate the suppression of fluctuations arising from exact global conservation of the conserved charges, we correct the susceptibilities using the subensemble acceptance method~\cite{Vovchenko:2020gne}. In this framework, only a fraction of the total system contributes to the measured susceptibilities, leading to a suppression in the higher-order cumulants as compared to the grand-canonical values. The acceptance factor, $\alpha$, is given as $\alpha = N_{\rm{ch}}(\Delta p_{\rm{T}}\Delta y)/N_{\rm{ch}}(\infty)$~\cite{Vovchenko:2020gne, Vovchenko:2021gas}. It modifies the susceptibilities as,
\begin{align}
\frac{\kappa_{2}}{\kappa_{1}} &= \beta\,\frac{\chi_{2}}{\chi_{1}}, \\[6pt]
\frac{\kappa_{3}}{\kappa_{2}} &= \frac{(1 - 2\alpha)\, \chi_{3}}{\chi_{2}}, \\[6pt]
\frac{\kappa_{3}}{\kappa_{1}} &= \frac{\beta(1 - 2\alpha)\, \chi_{3}}{\chi_{1}}, \\[6pt]
\frac{\kappa_{4}}{\kappa_{3}} &=
\frac{ (1 - 3\alpha\beta)\, \chi_{4}
      - 3\alpha\beta\, \dfrac{\chi_{3}^{2}}{\chi_{2}} }
     { (1 - 2\alpha)\, \chi_{3} }, \\[6pt]
\frac{\kappa_{4}}{\kappa_{2}} &=
\frac{ (1 - 3\alpha\beta)\, \chi_{4}
      - 3\alpha\beta\, \dfrac{\chi_{3}^{2}}{\chi_{2}} }
     { \chi_{2} }.
\end{align}
where, $\beta = (1 - \alpha)$.
\section{Results}
\label{Results}
\begin{figure*}
    \centering
    \includegraphics[width=0.45\linewidth]{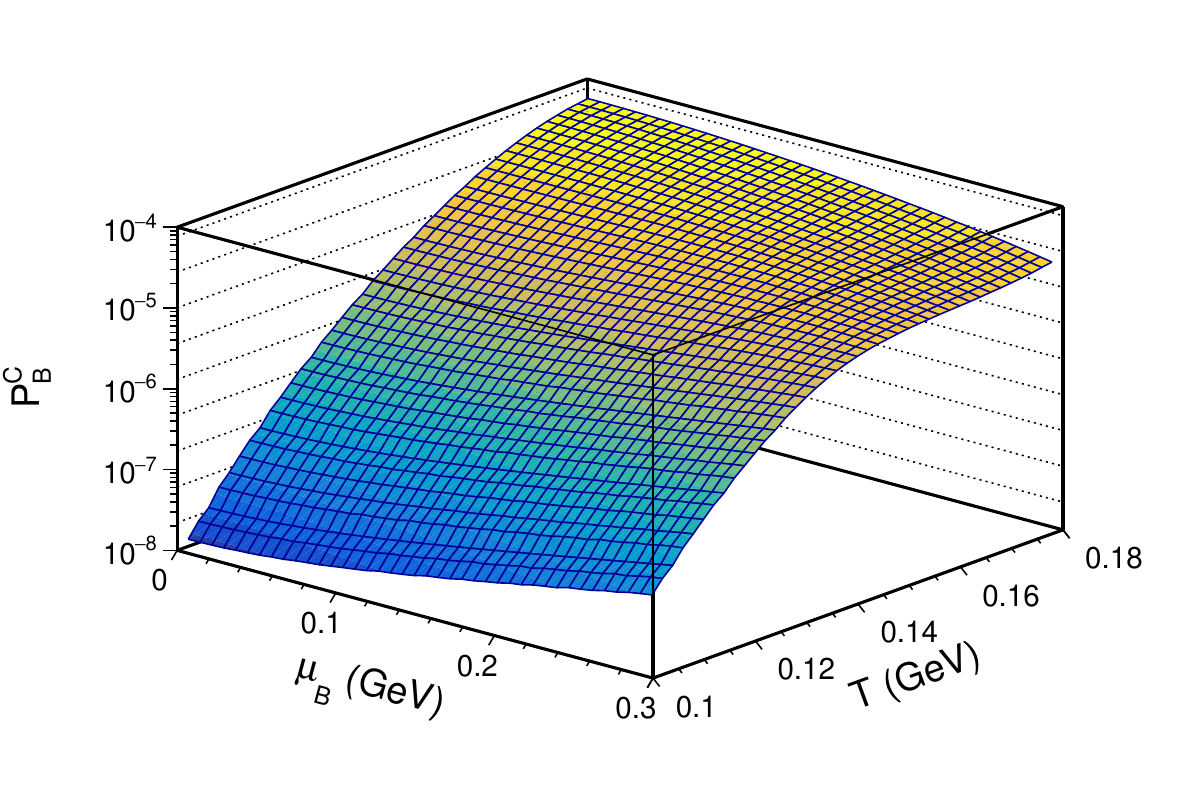}
    \includegraphics[width=0.45\linewidth]{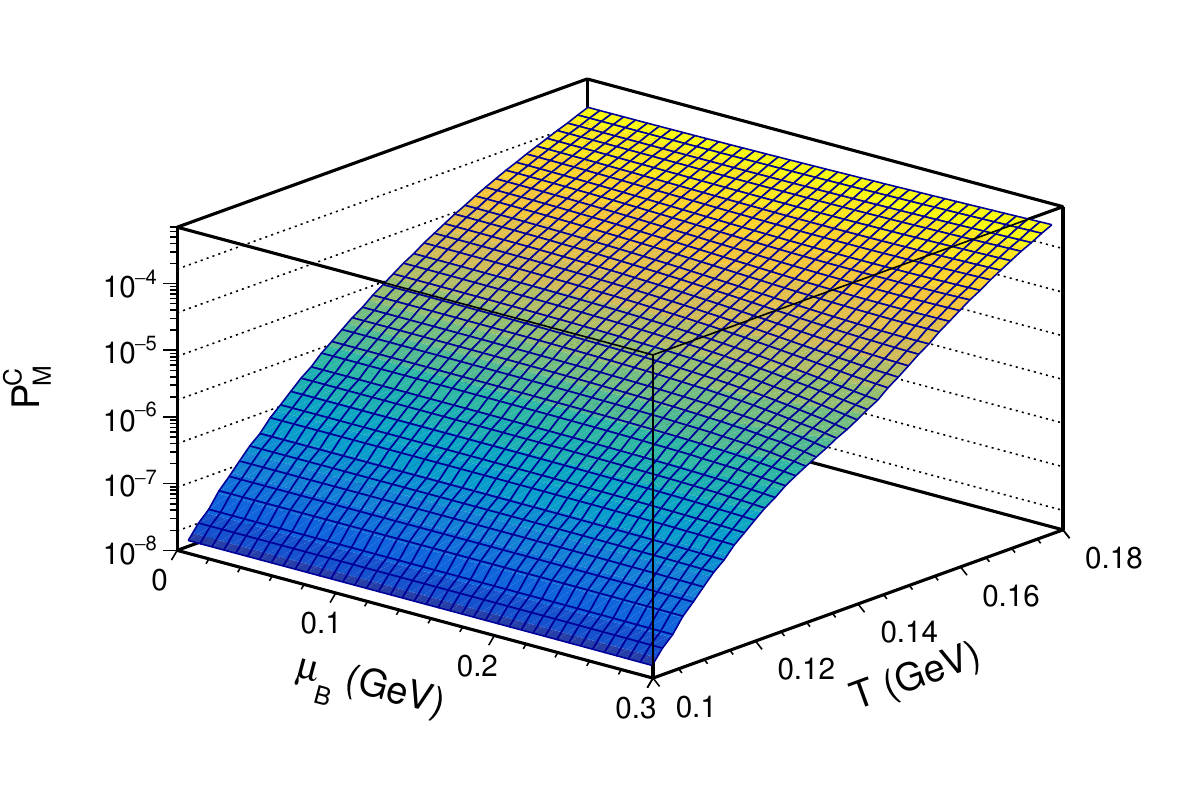}
    \caption{Partial pressure of charmed baryon~(left panel) and charmed meson~(right panel) as a function of chemical potential and temperature.}
    \label{fig_3D}
\end{figure*}

\begin{figure*}
    \centering
    \includegraphics[width=0.32\linewidth]{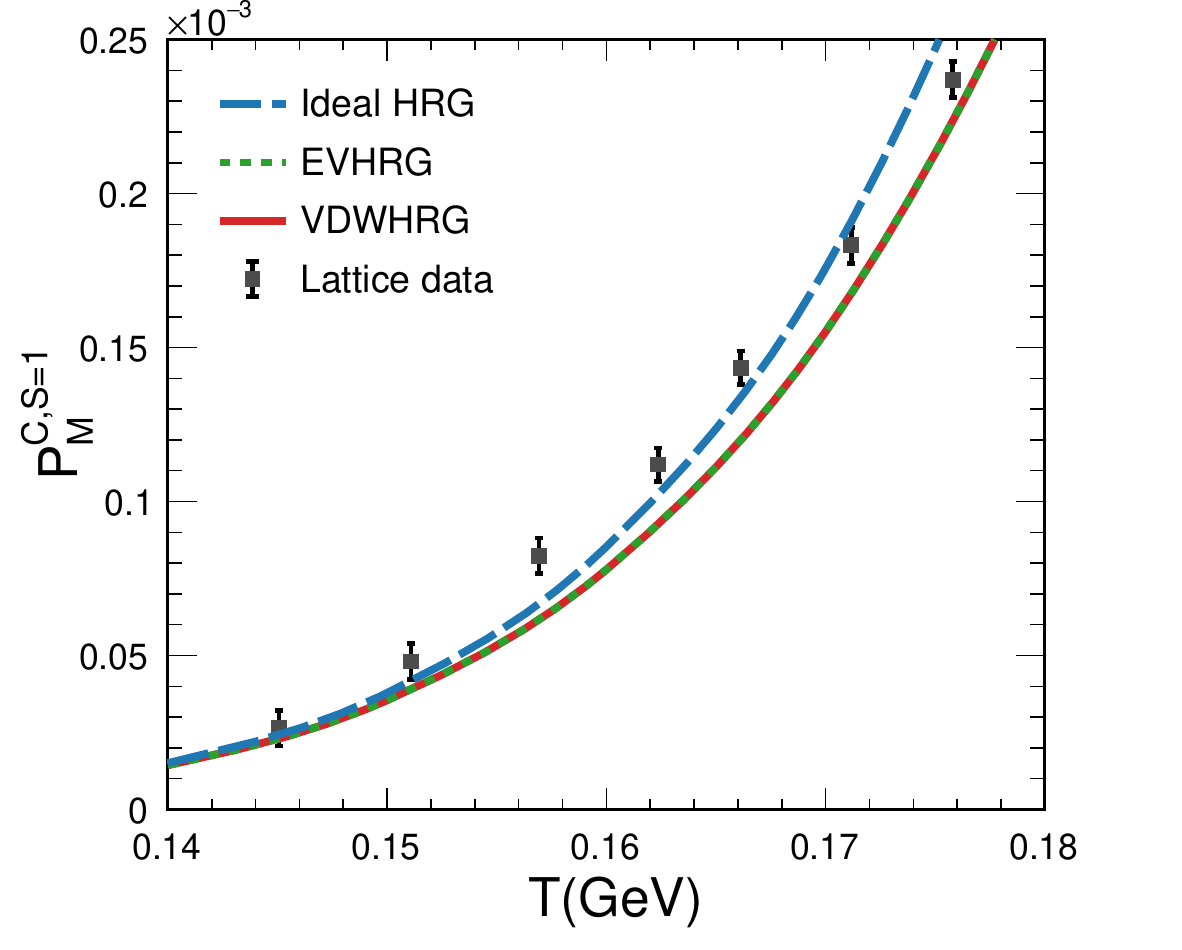}
    \includegraphics[width=0.32\linewidth]{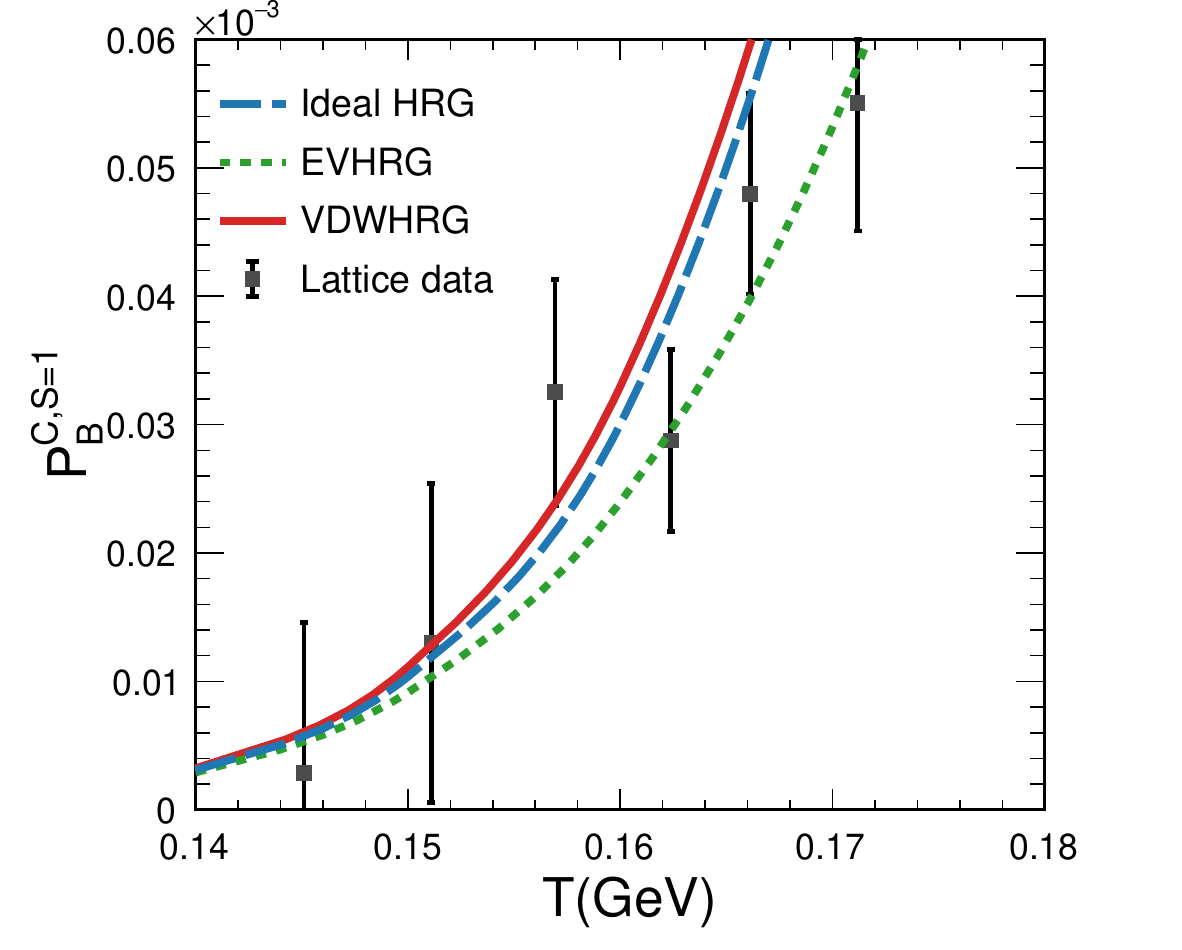}
    \includegraphics[width=0.32\linewidth]{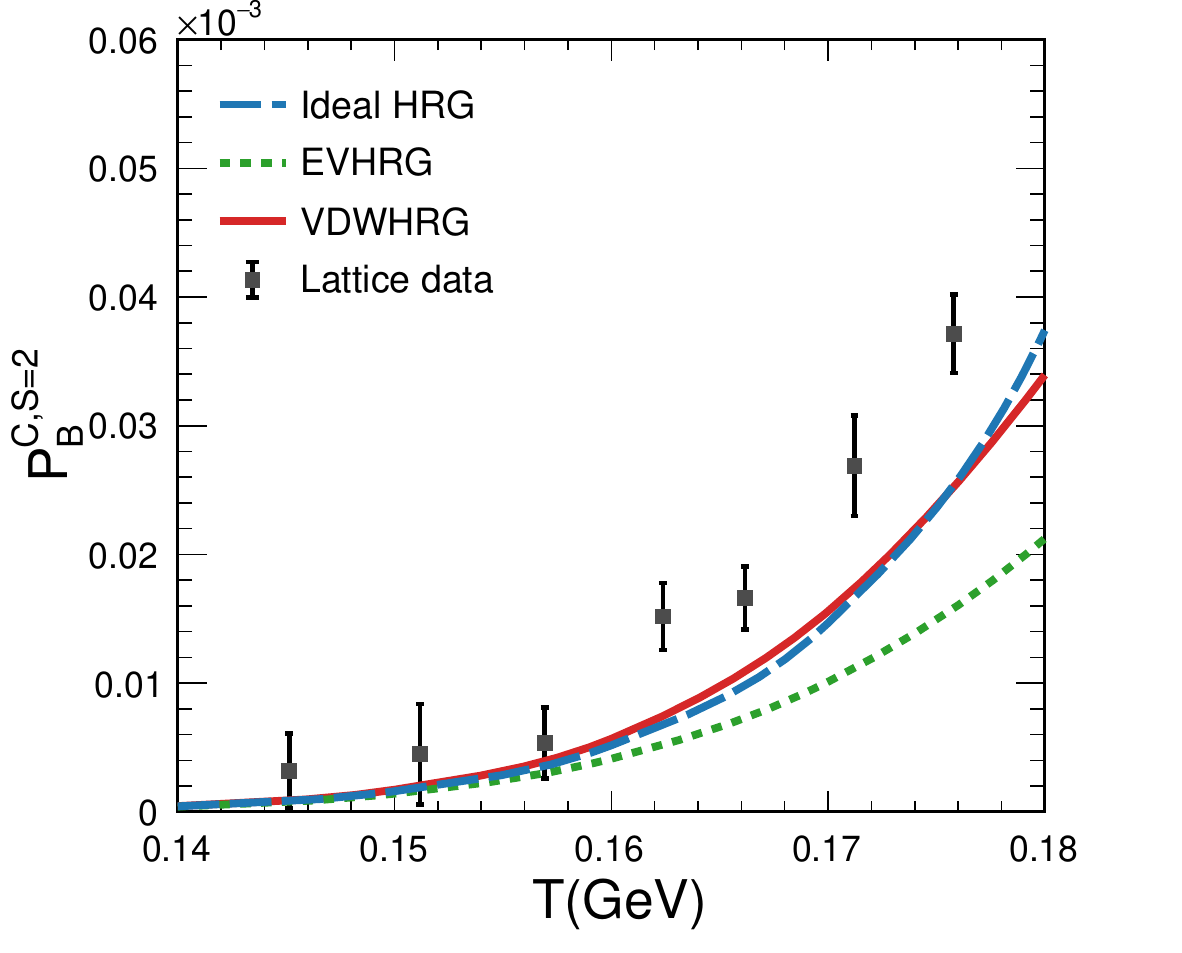}
    \caption{Partial pressure of charmed meson with strangeness = 1~(left), charmed baryon with strangeness = 1~(middle), and strangeness = 2~(right) as a function of temperature at $\mu_{\rm{B}}=0$. The markers are for lattice QCD results~\cite{Kaczmarek:2025dqt}.}
    \label{fig3}
\end{figure*}
For this work, we have used all the charmed hadrons in the PDG~\cite{ParticleDataGroup:2024cfk}, as well as the undiscovered charmed states from the quark model~\cite{Ebert:2009ua, Ebert:2011kk}. In the left panel of Fig.~\ref{fig1}, we present the partial pressure of charmed baryons, estimated using the quantity $(3\chi_{22}^{BC} - \chi_{13}^{BC})/2$. While the HRG model, including contributions from undiscovered states, overestimates this quantity and deviates from lQCD data for $T \gtrsim 160$ MeV, the VDWHRG model provides an excellent description up to $T \sim 180$ MeV. This suggests that the inclusion of attractive and repulsive interactions between hadrons, as implemented in the VDWHRG model, is crucial for describing the thermodynamics of charmed baryons near the crossover region. Similarly, in the right panel of Fig.~\ref{fig1}, we use Eq.~\ref{eq:PC_M} to estimate the partial pressure of charmed mesons as a function of temperature. We observe an increasing trend in the lattice QCD estimations throughout the entire temperature range, which is well described by all the HRG models. However, we still notice that incorporating interactions among the hadrons decreases the partial pressure of the charmed mesons. Moreover, in the left panel of Fig.~\ref{fig_3D}, we plot the partial pressure of the charmed baryon as a function of temperature and baryo-chemical potential. We observe that along the $\mu_{\text{B}}$ axis, the partial pressure of charmed baryon increases at low temperature, but drops slightly for higher $\mu_{\rm{B}}$ at higher temperatures. Additionally, in the right panel, we observe a monotonically increasing trend for the partial pressure of charmed mesons with increasing temperature. In comparison, with increasing baryo-chemical potential, the partial pressure of charmed mesons is unaffected.
\begin{figure*}
    \centering
    \includegraphics[width=0.32\linewidth]{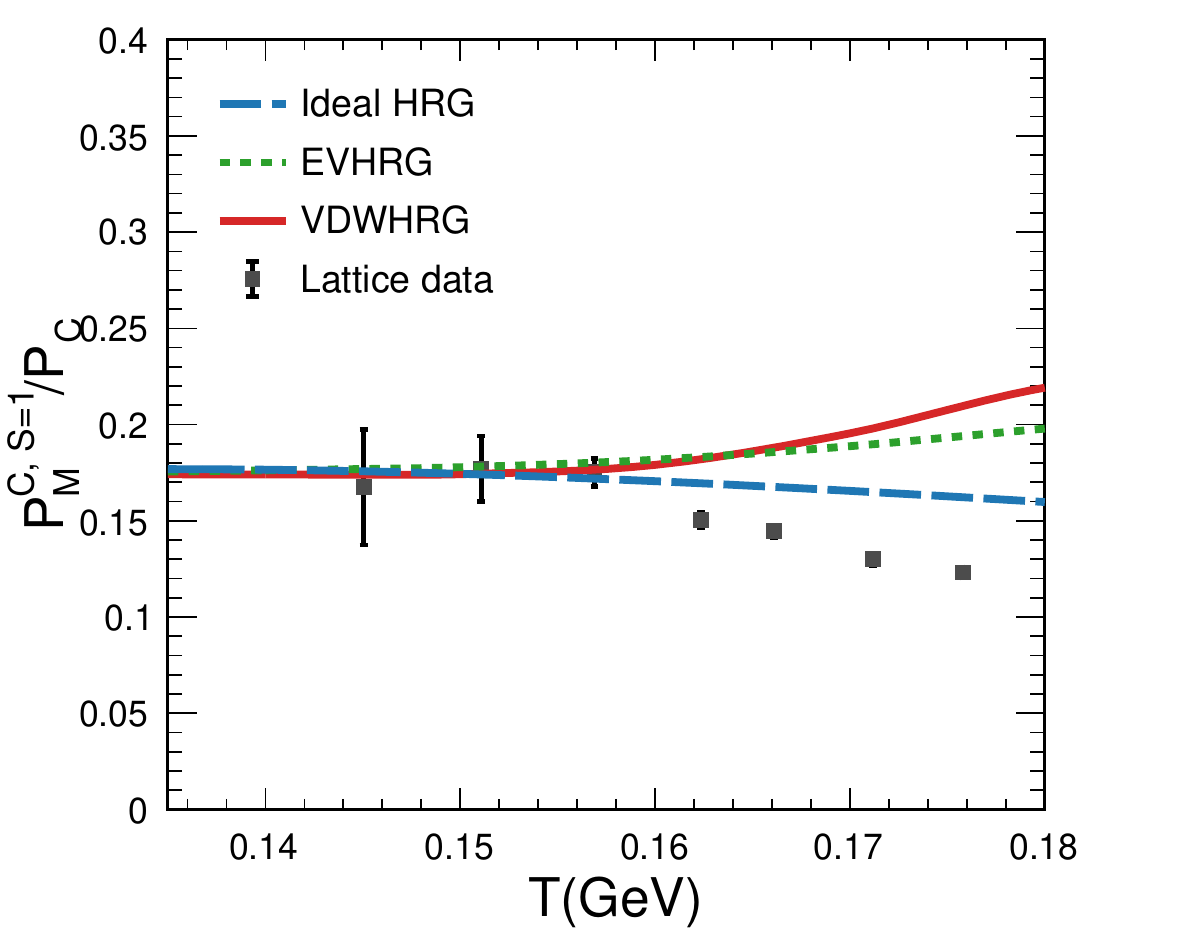}
    \includegraphics[width=0.32\linewidth]{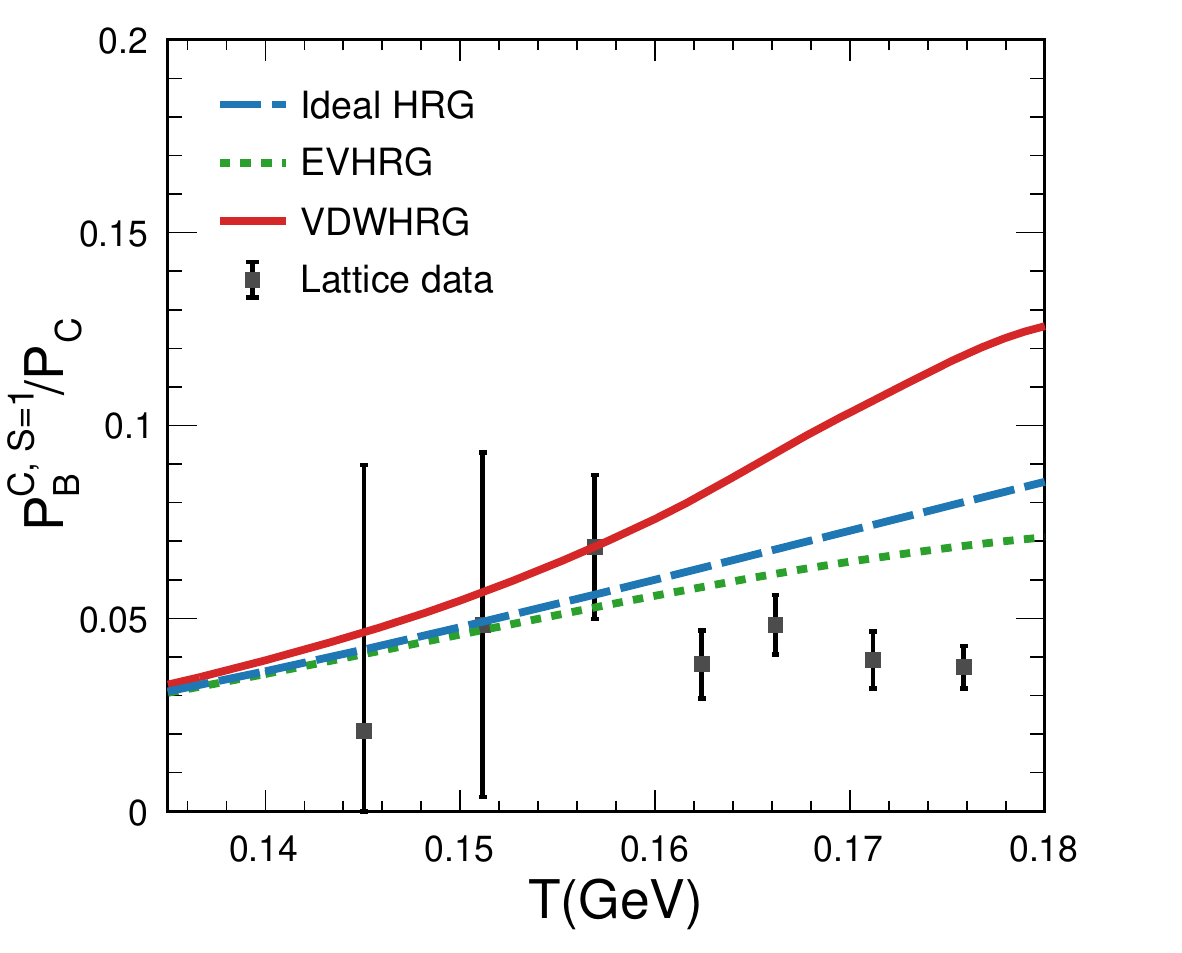}
    \includegraphics[width=0.32\linewidth]{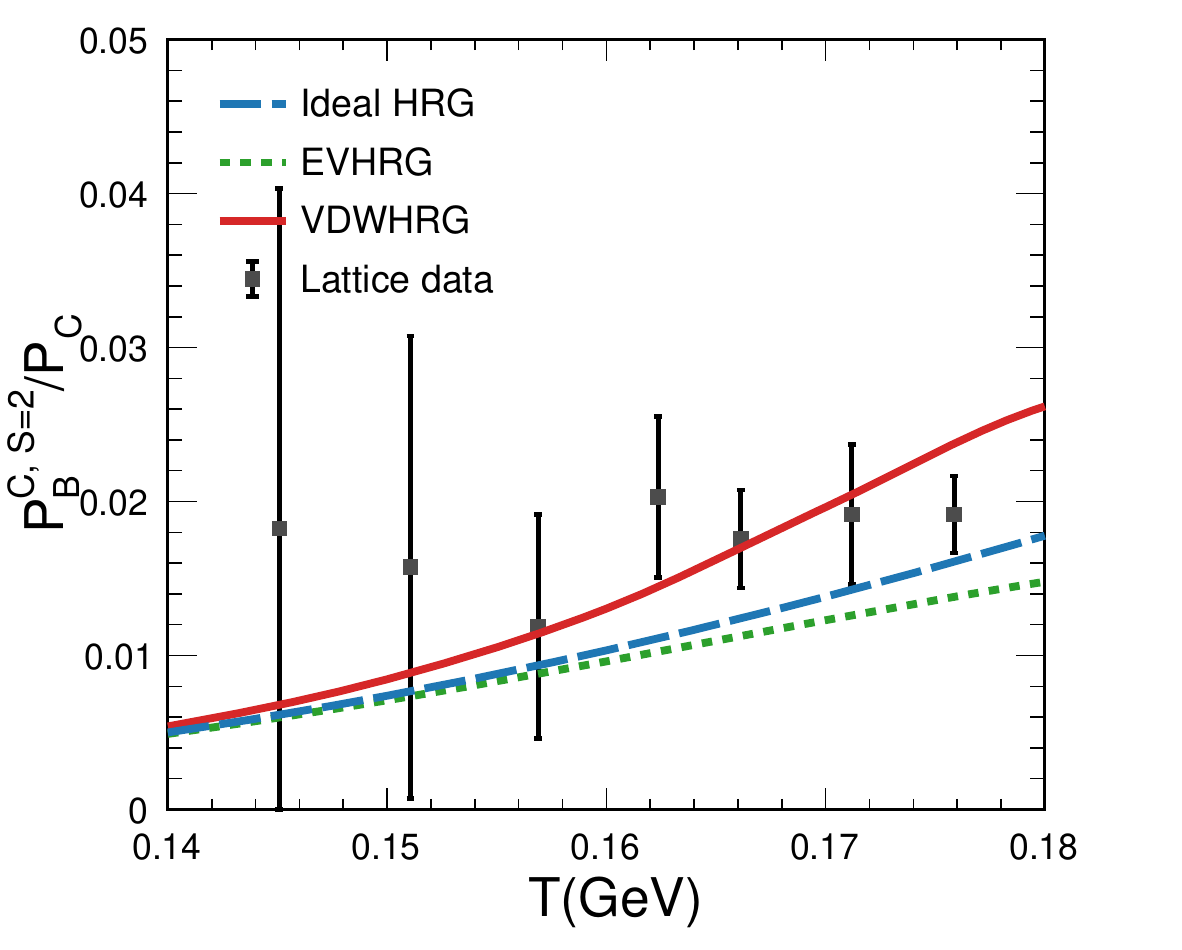}
    \caption{Partial pressure ratio at $\mu_{\rm{B}}=0$: Numerator: Charmed meson with strangeness = 1~(left), charmed baryons with strangeness = 1~(middle), and charmed baryons with strangeness = 2~(right). Denominator: Total charm pressure (meson + baryon), as a function of temperature. Compared with lattice QCD results from Ref.~\cite{Sharma:2024edf}.}
    \label{fig4}
\end{figure*}
\begin{figure}
    \centering
    \includegraphics[width=0.95\linewidth]{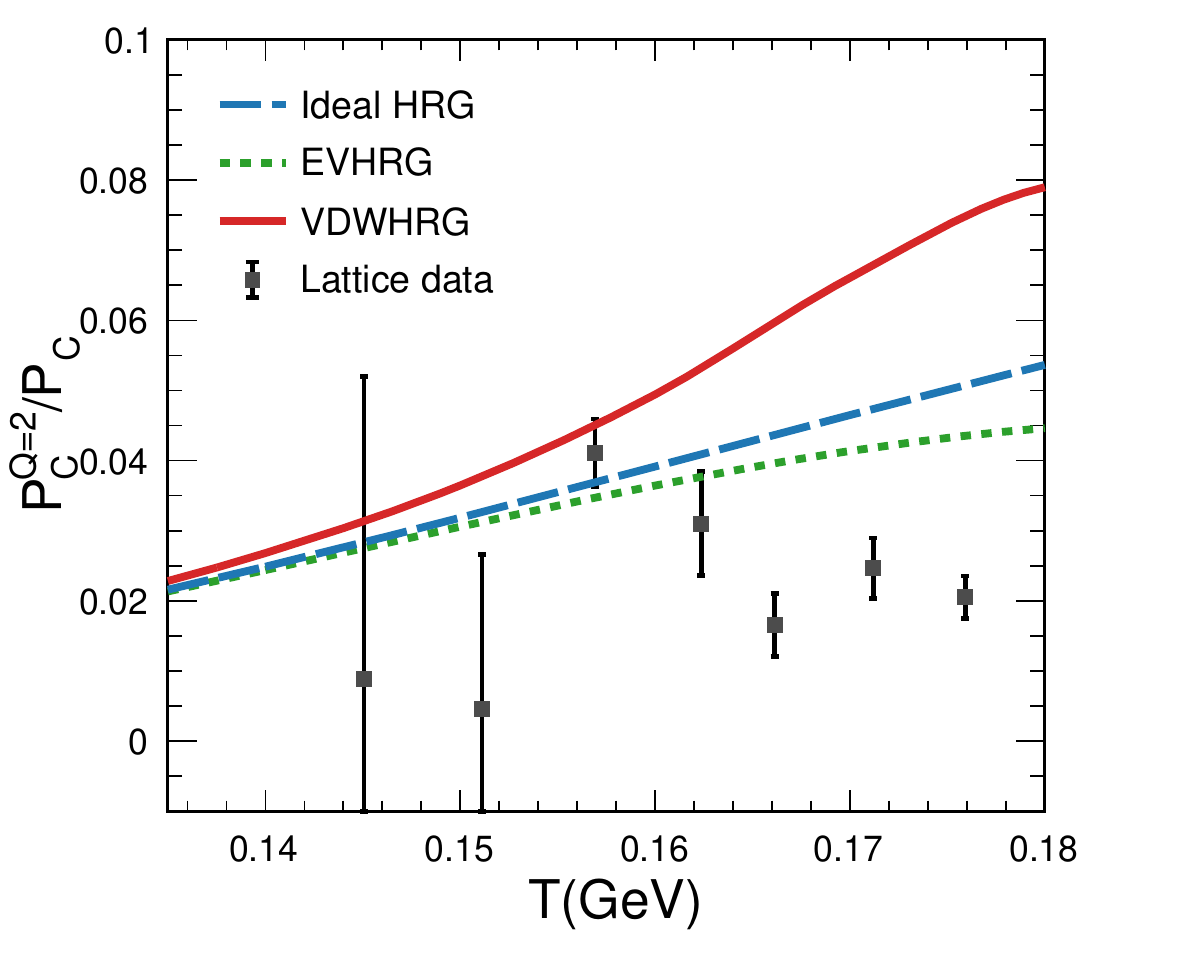}
    \caption{Partial pressure ratio at $\mu_{\rm{B}}=0$: Numerator: Charmed hadron with electric charge = 2. Denominator: Total charm pressure (meson + baryon), as a function of temperature compared with lattice QCD calculations~\cite{Sharma:2024edf}.}
    \label{fig5}
\end{figure}
Now, in the left panel of Fig.~\ref{fig3}, we study the partial pressure of charmed mesons with strangeness one. The IHRG, EVHRG, and VDWHRG models describe the lQCD data reasonably well, though the EVHRG and VDWHRG models provide a better match at high temperatures. The relatively good performance of the models for these states may be attributed to the fact that most of the expected charm-strange mesons (e.g., $D_s$ mesons) have been experimentally observed, allowing for a more complete implementation in the hadron resonance gas framework.
In the middle panel of Fig.~\ref {fig3}, we plot the partial pressure of charmed baryons with strangeness one. In this case, all the HRG models match the lQCD data within the large error bars. Although the EVHRG seems to show a reasonable agreement with the lattice QCD data at higher temperatures. Similarly, the right panel of Fig.~\ref{fig3} shows charmed baryons with strangeness two, and both the IHRG and VDWHRG models give similar values, which underpredict the lQCD estimations. In addition, the EVHRG shows even worse agreement with lQCD. This might be due to the fact that a larger number of charm-strange baryons, with $|S| = 2$, are still undiscovered and not predicted by the quark models yet. The similar values of IHRG and VDWHRG point to the fact that the effect of VDW is smaller as the number density of charmed baryons with strangeness two is very low, yet the attractive interactions counter the repulsive interactions reasonably. Future experimental efforts at facilities like LHCb and the upcoming Electron-Ion Collider (EIC) could help constrain these states better.

In Fig.~\ref{fig4} (left panel), we plot the ratio of partial pressure from charmed mesons with strangeness one to total charmed pressure. For the hadron gas case, we have used the total charm pressure as the added pressure from charmed mesons and charmed baryons. However, for the lQCD, they have also added a charm quark pressure, assuming a mixed state near the transition temperature. Due to this contribution from charm quark pressure, the lQCD trend goes down after $T \sim$ 160 MeV. The VDWHRG model explains the lQCD data up to $T\sim$~160 MeV, but then shows an upward trend after that. This mismatch comes from the neglected charm quark pressure, which seems to dominate after the transition temperature. 
\begin{figure*}
    \centering
    \includegraphics[width=0.45\linewidth]{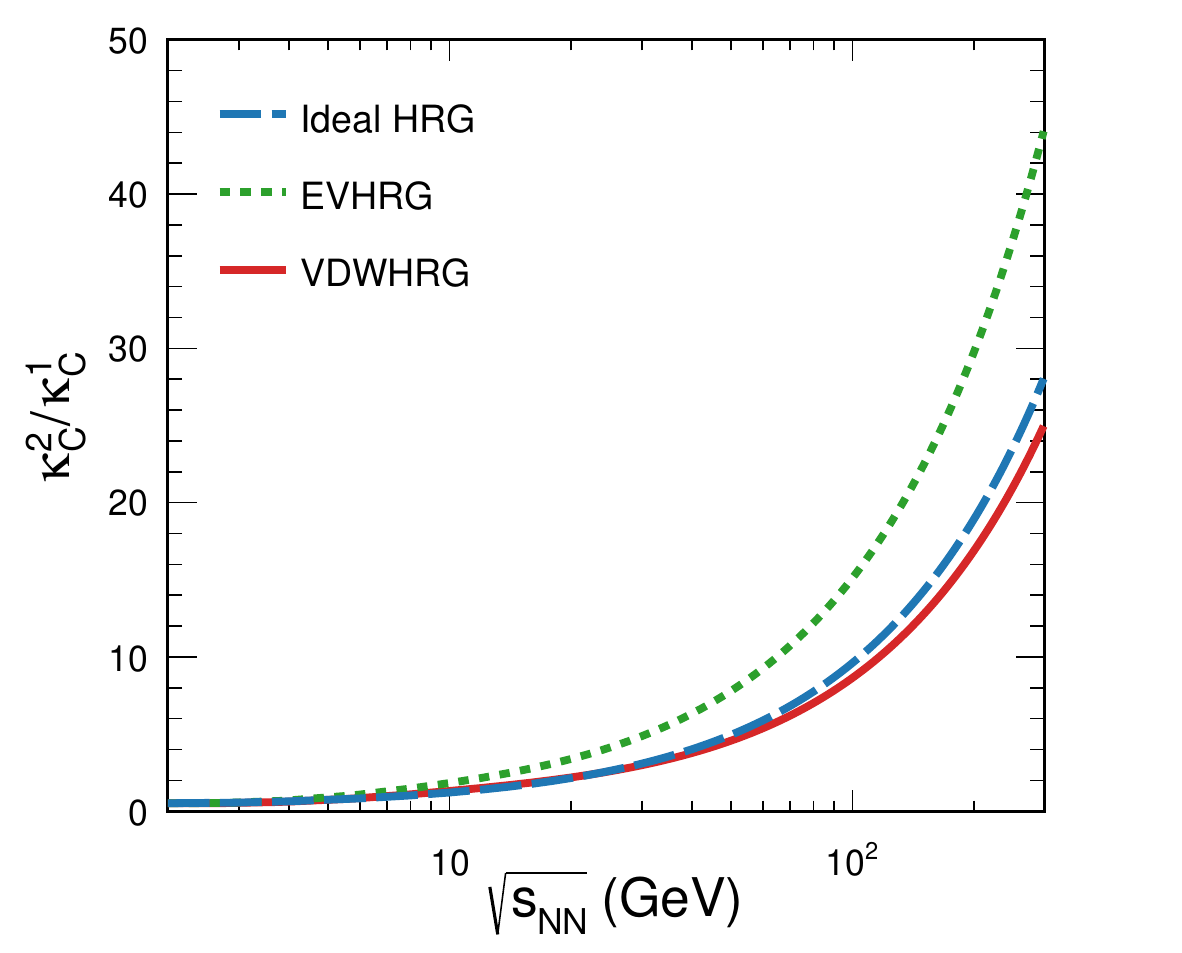}
    \includegraphics[width=0.45\linewidth]{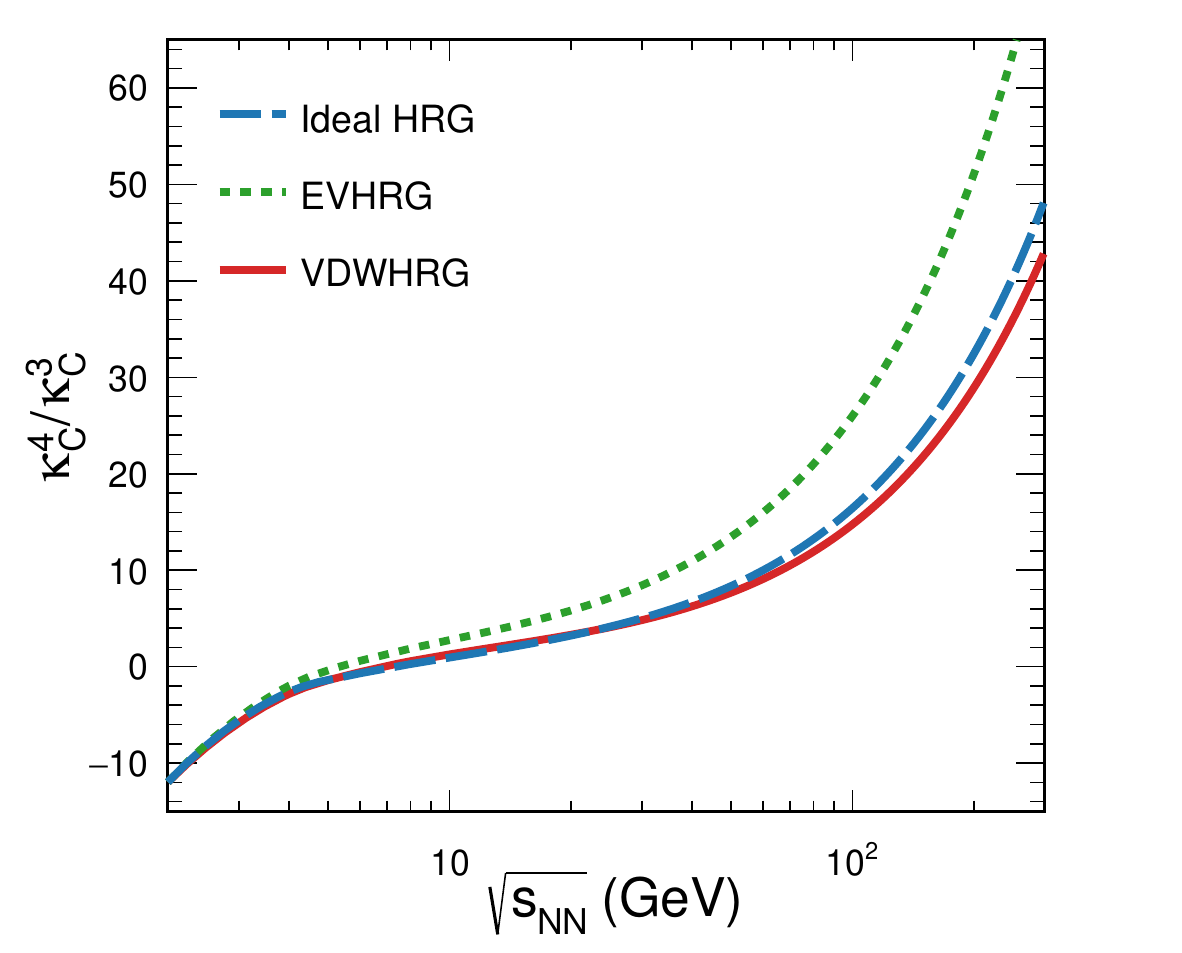}
    \caption{Charm susceptibilities ratios, $\kappa_2^C/\kappa_1^C$~(left) and $\kappa_4^C/\kappa_3^C$~(right), as a function of $\sqrt{s_{\rm NN}}$.}
        \label{fig6}
\end{figure*}

\begin{figure*}
    \centering
    \includegraphics[width=0.32\linewidth]{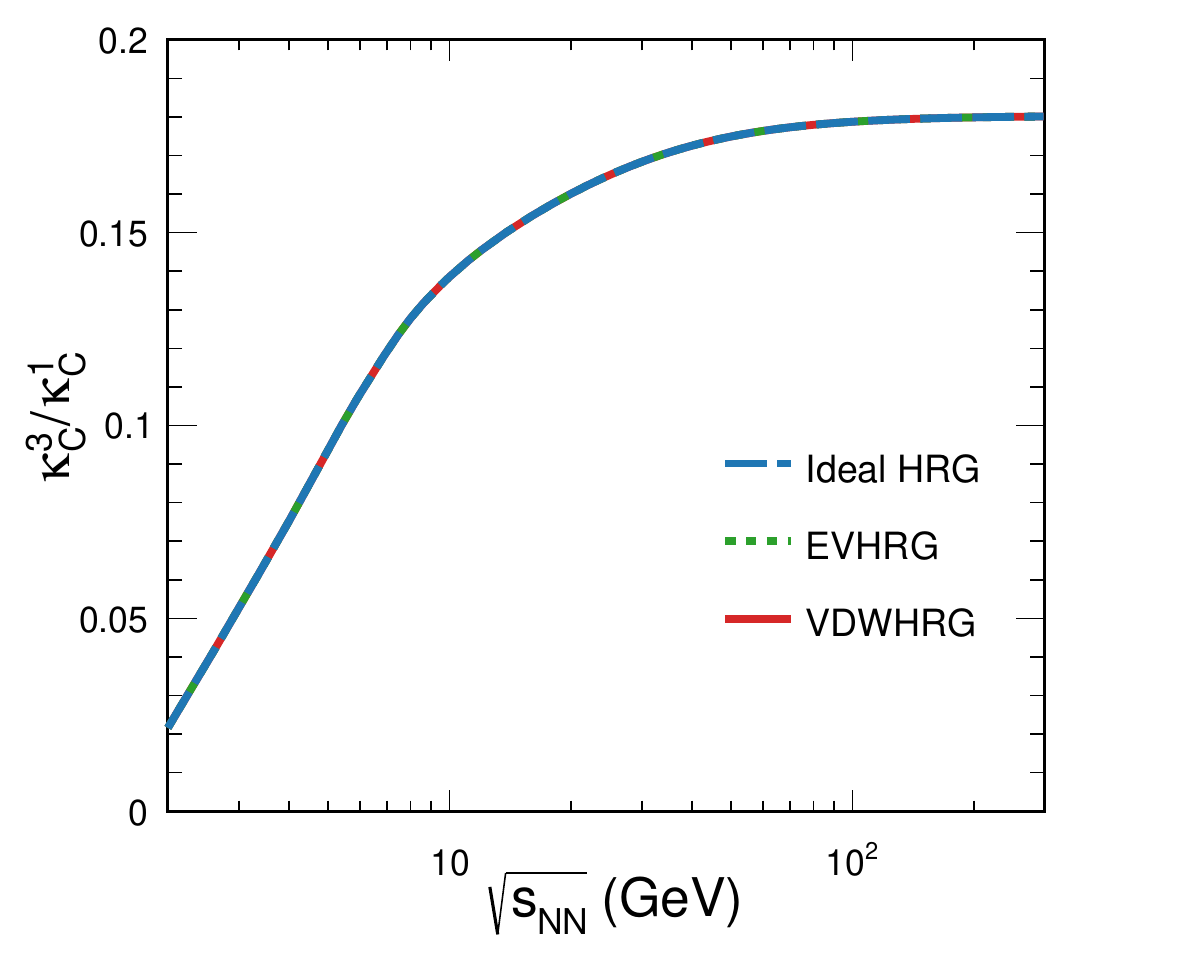}
     \includegraphics[width=0.32\linewidth]{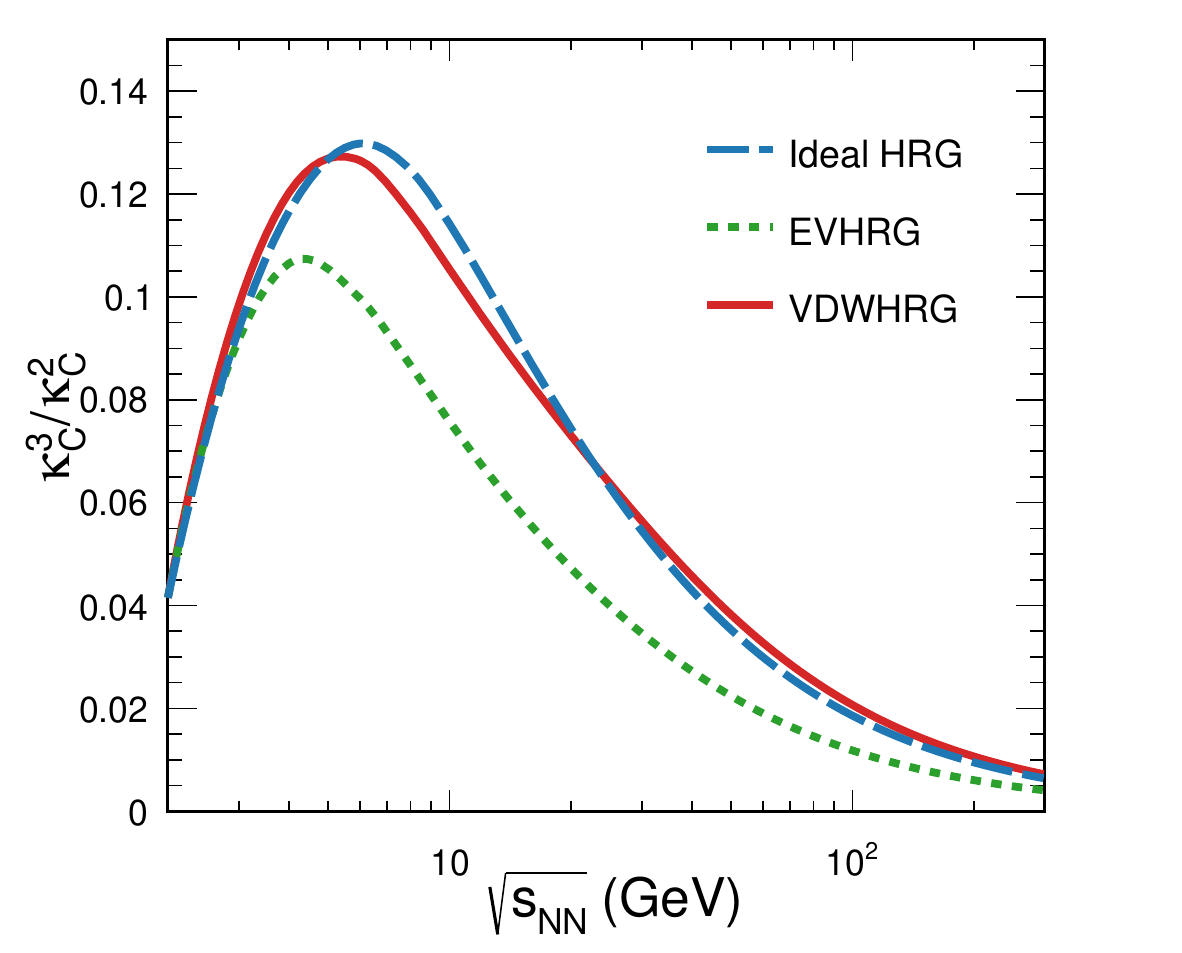}
    \includegraphics[width=0.32\linewidth]{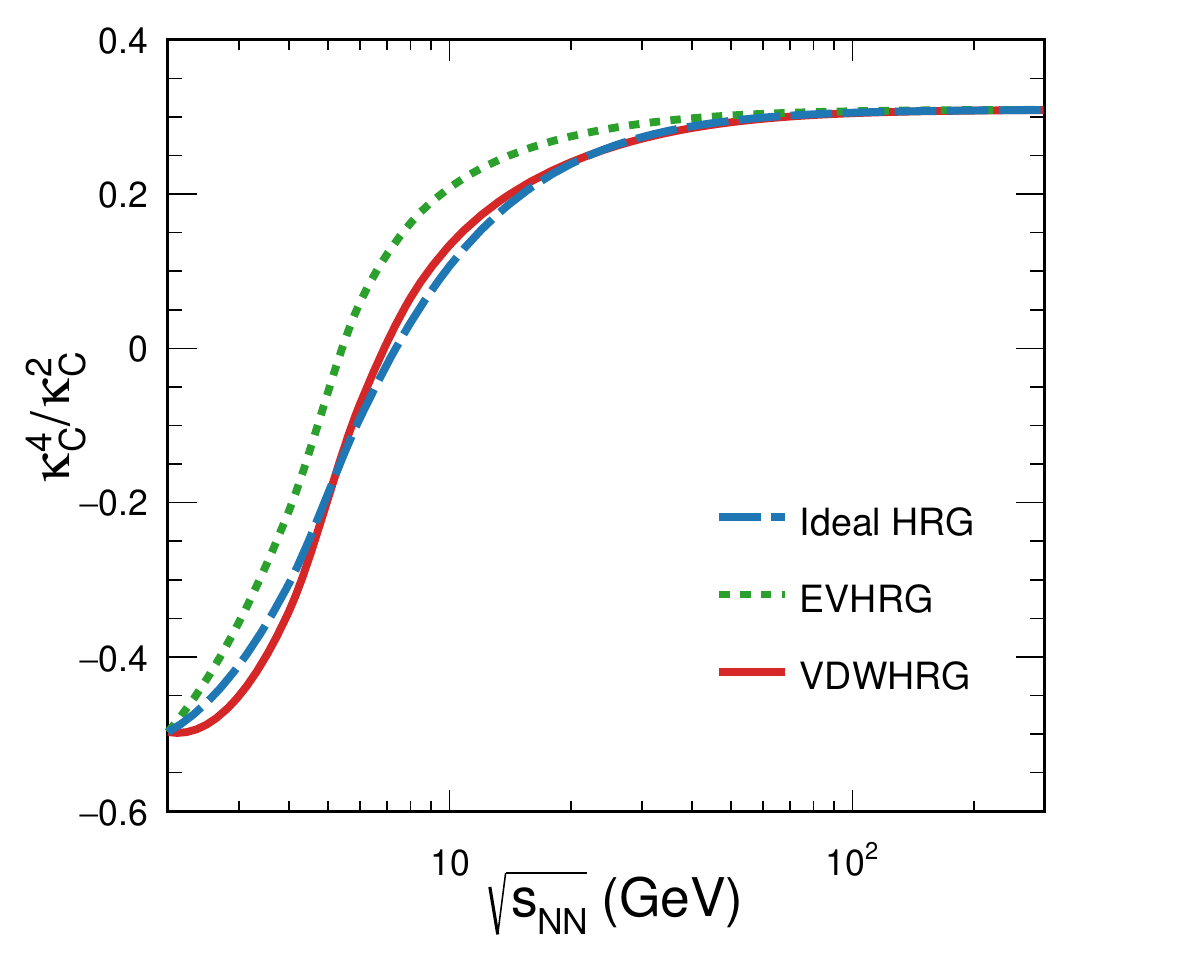}
    \caption{Charm susceptibilities ratios, $\kappa_3^C/\kappa_1^C$~(left), $\kappa_3^C/\kappa_2^C$~(middle), and $\kappa_4^C/\kappa_2^C$~(right), as a function of $\sqrt{s_{\rm NN}}$.}
      \label{fig7}
\end{figure*}
In the middle panel of Fig.~\ref{fig4}, we show the ratio of the partial pressure of charmed baryons with strangeness one to the total charm pressure. Here too, the VDWHRG shows a similar behavior as the lQCD data up to $ T\sim$160 MeV, after which a clear deviation can be seen, owing to the non-inclusion of charm quark pressure. Fig.~\ref{fig4} (right panel) shows the ratio of charmed baryon with strangeness 2 to total charm pressure as a function of temperature. For $T < 160$~MeV, the predictions of the VDWHRG model remain consistent with the lattice QCD results within their large uncertainties. However, for $T>160$~MeV, the lQCD estimation shows a saturation behaviour that none of the HRG models are able to reproduce.

Finally, we show the ratio of charged-charm partial pressure to total charm pressure in Fig.~\ref {fig5}. The charmed state for the numerator has charge $Q$ = 2. Similar to the previous plots in Fig.~\ref{fig4}, here also we see a deviation of VDWHRG to lQCD data after $T \sim$~160 MeV, which hints towards the contribution from the quark sector, suggesting that the transition from hadron gas to a deconfined medium occurs near this temperature. In all HRG models, these ratios estimated above serve as important probes of the thermal distribution of charm among different hadronic species. They provide insight into how charm quarks hadronize in the confined phase and how the charm-carrying degrees of freedom evolve with temperature. Comparing HRG model predictions with lQCD data for these ratios helps us to test the completeness of the hadron spectrum, and can indicate the presence of missing or unobserved charmed states. Additionally, these ratios are relevant for understanding charm production and freeze-out conditions in heavy-ion collisions.

On the other hand, the charm susceptibilities and their ratios as functions of collision energy reveal a multi-faceted picture of QCD matter across different phases. Much like in the baryon sector, in Fig.~\ref{fig6} the increasing trends of $\kappa_2^C/\kappa_1^C$ and $\kappa_4^C/\kappa_3^C$ with $\sqrt{s_{\rm NN}}$ demonstrate the gradual emergence of correlated charm production mechanisms, transitioning from independent fragmentation at low energies to medium-dominated processes in the QGP phase at high energies. The consistent model hierarchy ({VDWHRG $<$ IHRG $<$ EVHRG}) highlights how van der Waals interactions most effectively capture the partial thermalization of charm quarks, with attractive forces suppressing fluctuations closer to lQCD predictions. This is complemented by the constancy of $\kappa_3^C/\kappa_1^C$ and $\kappa_4^C/\kappa_2^C$ saturating beyond center-of-mass energy of 10--30 GeV as shown in the left and right panels of Fig.~\ref{fig7}, which serve as a thermodynamic fingerprint of charm's unique properties, showing that its large mass preserves the charm symmetry and Gaussian statistics despite the phase transition. However, at lower energies, the ratios show an increasing behavior before saturating beyond $\sqrt{s_{\rm{NN}}} = $ 10 GeV due to the consideration of global conservation of charges through the acceptance factor. During this increase, an effect of interaction among the hadrons can be seen in the higher-order cumulant ratio. The decreasing trend of $\kappa_3^C/\kappa_2^C$, beyond $\sqrt{s_{\rm{NN}}} = $ 10 GeV, as shown in the middle panel of Fig.~\ref{fig7} and its saturation at high energies reflects the decoupling of charm from baryon number in the $\mu_B \to 0$ regime. Here, for center-of-mass energies lower than 10 GeV, an interplay between the effects of global conservation and charm production is observed, which forces the ratio to exhibit an increasing behavior. This comprehensive view positions charm fluctuations as both a sensitive probe of the QCD phase diagram and a stringent test for theoretical models across the entire collision energy spectrum.

The systematic evolution of these ratios with collision energy offers a powerful tool for mapping the QCD phase diagram. Future measurements with improved precision could further constrain the location of the critical point and clarify the temperature dependence of charm hadronization. In addition to this, one key feature that could substantially change these ratios is the charm chemical potential. At RHIC energies, the charm chemical potential ($\mu_c$) can be finite due to the nature of charm quark production and conservation in heavy-ion collisions. Charm quarks are predominantly produced in initial hard scatterings rather than through thermal excitation, and their number remains approximately conserved throughout the evolution of the fireball. In thermal models like the HRG, a finite $\mu_c$ can be introduced to enforce this conservation, especially when using a grand canonical ensemble. Moreover, local imbalances between charm and anti-charm quarks can develop due to event-by-event fluctuations or limited charm diffusion, further justifying a nonzero $\mu_c$~\cite{Song:2024rjh}. In thermal fits to charm hadron yields (such as $D^0$, $\Lambda_c$, or $J/\psi$), incorporating a finite charm chemical potential is necessary to correctly reproduce observed yields under the constraint of fixed net charm. Thus, a thorough look into charm chemical potential and subsequent charm susceptibility estimation is very much timely and necessary.

\section{Summary}
\label{sum}
In this work, we study the thermodynamic properties and charm sector fluctuations of QCD matter using the van der Waals Hadron Resonance Gas (VDWHRG) model. By incorporating both attractive and repulsive interactions between hadrons, the VDWHRG model significantly improves agreement with lattice QCD (lQCD) data, especially in the temperature range up to 180~MeV. We show that the VDWHRG model provides an improved description of charm susceptibilities and partial pressures. Our study includes the estimation of partial pressures of charmed mesons and baryons, both with and without strangeness, and compares them to available lQCD results. While VDWHRG captures the qualitative trends well, especially below $T\sim160$~MeV, discrepancies at higher temperatures highlight the possible role of deconfined charm degrees of freedom not accounted for in hadron-based models. Additionally, we extend our analysis to baryon-rich regimes, offering insights into charm susceptibilities in regions not easily accessible by lQCD. Furthermore, we study the charm susceptibilities and their ratio as a function of collision energy, which highlights their sensitivity to the underlying QCD phase structure. The observed trend for $\kappa_{2}^{C}/\kappa_{1}^{C}$ and $\kappa_{4}^{C}/\kappa_{3}^{C}$ shows the gradual increase in charm production with an increase in collision energy. Moreover, a saturation behavior in $\kappa_{3}^{C}/\kappa_{2}^{C}$ highlights the decoupling of the charm and baryon number at higher center-of-mass collision energy. For the study of charm susceptibility as a function of center-of-mass energy, we make use of the parametrization established from the light hadron yield. Even though, for the scope of our work, it is sufficient, in the future, one can determine a charm-specific freeze-out condition, for instance, through a statistical hadronization model fit to charm hadron ratios. The upcoming high-luminosity upgrades at the LHC and higher-energy collisions are expected to provide sufficient data for the charm sector to work on such a parametrization.

To summarize, the results presented here provide a strong motivation for further constraining charm-sector interactions through both theoretical and experimental efforts. Theoretically, our predictions for partial pressure at finite $\mu_{B}$ can be tested by future lattice QCD calculations that employ novel methods to circumvent the sign problem. Furthermore, comparing the VDWHRG approach with other models and effective field theories could offer deeper insights into the microscopic origin of these interactions. Experimentally, the most direct path to constrain the VDW parameters is through the measurement of open-charm fluctuations and correlations in heavy-ion collisions, which remains a novel direction and has not been thoroughly explored. Higher-order moments of the distributions of D mesons and $\Lambda_{c}$ baryons are directly sensitive to the charm susceptibilities discussed in this work. Additionally, correlation measurements, such as $D-p$ and $D-\Lambda$ femtoscopy, can provide crucial independent constraints on the scattering parameters and effective forces between charm and light hadrons. The upcoming high-statistics data from the LHC Run 3 and the Beam Energy Scan II at RHIC, as well as future programs at FAIR and NICA, will be instrumental in performing these decisive tests, thereby transforming the qualitative success of the VDWHRG model into a quantitative extraction of charm-hadron interactions in the hot and dense medium.

\section*{Acknowledgments}
K.G. acknowledges financial support from the Prime Minister's Research Fellowship (PMRF), Government of India. K.K.P. gratefully acknowledges the financial aid from the University Grants Commission (UGC), Government of India. The authors also acknowledge the DAE-DST, Government of India, funding under the mega-science project “Indian participation in the ALICE experiment at CERN” bearing Project No. SR/MF/PS-02/2021-IITI(E-37123).

\newpage
\onecolumngrid

\appendix

\section{Partial pressure for charmed baryons}
\label{charmedP}

We begin by writing the pressure in a Taylor series expansion around vanishing chemical potentials $(\mu_B, \mu_Q, \mu_S, \mu_C) = 0$. This allows us to express $P/T^4$ as a sum over derivatives of the pressure with respect to the reduced chemical potentials $\hat{\mu}_{a} = \mu_{a} / T$,

\begin{equation}
\frac{P}{T^4} = \sum_{k,l,m,n} \frac{1}{k!~l!~m!~n!}
\frac{\partial^{k+l+m+n}\left[P(\mu_{B}, \mu_{Q}, \mu_{S}, \mu_{C})/T^{4}\right]}
{\partial \hat{\mu}_{B}^{k} \, \partial \hat{\mu}_{Q}^{l} \, \partial \hat{\mu}_{S}^{m} \, \partial \hat{\mu}_{C}^{n}} 
\bigg |_{\mu = 0} \,
\hat{\mu}_{B}^{k} \, \hat{\mu}_{Q}^{l} \, \hat{\mu}_{S}^{m} \, \hat{\mu}_{C}^{n}.
\end{equation}

Here, $k$, $l$, $m$, and $n$ are non-negative integers representing the order of the derivative with respect to each chemical potential.  

For convenience, we introduce generalized susceptibilities $\chi^{BQSC}_{klmn}$, which compactly represent the derivatives appearing above. The expansion then reads,

\begin{equation}
\frac{P}{T^4} = \sum_{k,l,m,n} \frac{1}{k!~l!~m!~n!}
\chi^{BQSC}_{klmn}~\hat{\mu}_{B}^{k}~\hat{\mu}_{Q}^{l}~\hat{\mu}_{S}^{m}~\hat{\mu}_{C}^{n}.
\label{PC1}
\end{equation}

The generalized susceptibility is given as, 
\begin{equation}
\label{equation_susceptibility_appdx}
\chi^{BQSC}_{klmn} = \frac{\partial^{k+l+m+n}[P(\mu_{B}, \mu_{Q}, \mu_{S}, \mu_{C})/T^{4}]}{\partial \hat{\mu}_{B}^{k} ~\partial \hat{\mu}_{Q}^{l} ~\partial \hat{\mu}_{S}^{m} ~\partial \hat{\mu}_{C}^{n}}.
\end{equation}

Now, the total charm contribution to the pressure can be written as~\cite{Bazavov:2014yba, Kaczmarek:2025dqt, Sharma:2024edf},
\begin{equation}
    \frac{P_{C}(T,\vec{\mu})}{T^{4}} = M_{C}(T,\vec{\mu}) + B_{C}(T,\vec{\mu}), 
\end{equation}
where $M_{C}(T,\vec{\mu})$ and $B_{C}(T,\vec{\mu})$ are the contributions from charmed mesons and baryons, respectively. Since the charmed hadrons are significantly heavier than the temperature range relevant for studying thermodynamics near the QCD crossover, the Boltzmann approximation is suitable for all charmed states. $M_{C}(T,\vec{\mu})$ and $B_{C}(T,\vec{\mu})$ is given as~\cite{Bazavov:2014yba, Kaczmarek:2025dqt, Sharma:2024edf},
\begin{equation}
    M_{C}(T,\vec{\mu}) = \frac{1}{2\pi^2} 
    \sum_{i} g_i \left( \frac{m_i}{T} \right)^2 
    K_2(m_i/T) \, \cosh\big(Q_i\hat{\mu}_Q 
    + S_i\hat{\mu}_S + C_i\hat{\mu}_C \big).
    \label{PC_meson}
\end{equation}

\begin{equation}
    B_{C}(T,\vec{\mu}) = \frac{1}{2\pi^2} 
    \sum_{i} g_i \left( \frac{m_i}{T} \right)^2 
    K_2(m_i/T) \, \cosh\big( B_i\hat{\mu}_B + Q_i\hat{\mu}_Q 
    + S_i\hat{\mu}_S + C_i\hat{\mu}_C \big).
    \label{PC_baryon}
\end{equation}
Here, $g_i$ is the degeneracy factor, $m_i$ is the hadron mass, and $K_2$ is the modified Bessel function of the second kind. For charmed baryons, we can rewrite Eq.~\ref{PC_baryon}, and we expand the $\cosh{\rm{X}}$ terms up to 4th order,
\begin{equation}
    \frac{P}{T^{4}} = P_{B}^{C}\cosh{\rm{X}} = P_{B}^{C} \left[ 1 + \frac{X^{2}}{2!} + \frac{X^{4}}{4!} + \cdots \right].,
    \label{PC2}
\end{equation}
where, $P_{B}^{C}$ and $\rm{X}$ are given as,
\begin{equation*}
    P_{B}^{C} = \frac{1}{2\pi^2} \sum_{i} g_i \left( \frac{m_i}{T} \right)^{2}K_2(m_i/T),
\end{equation*}
\begin{equation*}
    \rm{X} = B_i\hat{\mu}_{B} + Q_i\hat{\mu}_{Q} + S_i\hat{\mu}_{S} + C_i\hat{\mu}_{C}.
\end{equation*}

The fourth-order term in this expansion is,
\begin{equation}
    \frac{P}{T^{4}} = P_{B}^{C}~\frac{X^{4}}{4!}
    \label{p_alpha}
\end{equation}

Expanding $X^{4}$ using the multinomial theorem gives us,
\begin{equation}
    {\rm{X^{4}}} = \sum_{k+l+m+n = 4}\frac{4!}{k!~l!~m!~n!}B^{k}Q^{l}S^{m}C^{n}~\hat{\mu}_{B}^{k}~\hat{\mu}_{Q}^{l}~\hat{\mu}_{S}^{m}~\hat{\mu}_{C}^{n}
    \label{X_exp}
\end{equation}

We now compare Eq.~\ref{PC1} and Eq.~\ref{X_exp} by matching coefficients of specific monomials in $\hat{\mu}_B, \hat{\mu}_Q, \hat{\mu}_S, \hat{\mu}_C$.  To begin with, we consider the coefficient of $\hat{\mu}^{2}_{B}~\hat{\mu}^{2}_{C}$. From Eq.~\ref{PC1}, the coefficient of $\hat{\mu}^{2}_{B}~\hat{\mu}^{2}_{C}$ is,
\begin{equation}
    \frac{1}{2!~2!}~\chi_{22}^{BC}
    \label{PC1_coeff}
\end{equation}

Combining Eq.~\ref{p_alpha} and Eq.~\ref{X_exp}, the coefficient of $\hat{\mu}^{2}_{B}~\hat{\mu}^{2}_{C}$ for baryons (where $B=1$ and $C=1$) is:
\begin{equation}
    P_{B}^{C}~\frac{1}{4!}\frac{4!}{2!~2!}
    \label{PC2_coeff}
\end{equation}

Equating Eq.~\ref{PC1_coeff} and Eq.~\ref{PC2_coeff}, we can write,
\[
\chi_{22}^{BC} = P_{B}^{C}.
\]
Similarly, considering the coefficient of $\hat{\mu}^{1}_{B}~\hat{\mu}^{3}_{C}$ gives:
\[
\chi_{13}^{BC} = P_{B}^{C}.
\]
From Eq.~\ref{PC_meson}, one can see that for meson we have $B=0$, and as a result the $\hat{\mu}^{2}_{B}~\hat{\mu}^{2}_{C}$ and $\hat{\mu}^{1}_{B}~\hat{\mu}^{3}_{C}$ term do not arise in the expansion. Now, the pressure arising in the system due to the charm quarks is given as~\cite{Sharma:2024edf},
\begin{equation}
    \frac{P}{T^{4}} = \frac{3}{\pi^{2}} \left( \frac{m_{C}}{T} \right )^{2}K_{2}\left( \frac{m_{C}}{T} \right ) \cosh\big( \frac{1}{3}\hat{\mu}_B + \frac{2}{3}\hat{\mu}_Q + \hat{\mu}_C \big)
\end{equation}
In a similar fashion, we define, 
\begin{equation*}
    P^{C}_{q} =  \frac{3}{\pi^{2}} \left( \frac{m_{C}}{T} \right )^{2}K_{2}\left( \frac{m_{C}}{T} \right )
\end{equation*}
Following the similar method, we take  $B=\frac{1}{3}$. On recomputing Eq.~\ref{PC2_coeff} for charm quarks, the coefficient of $\hat{\mu}^{2}_{B}~\hat{\mu}^{2}_{C}$ is,
\begin{equation}
        \frac{1}{4!}\frac{4!}{2!~2!} \left( \frac{1}{3} \right)^{2} P^{C}_{q}
    \label{PC2_coeff_quarks}
\end{equation}
However, the coefficient of $\hat{\mu}^{2}_{B}~\hat{\mu}^{2}_{C}$ from Eq.~\ref{PC1} remains same, which gives us,
\begin{equation}
    \chi_{22}^{BC} = P^{C}_{q}/9.
\end{equation}
Similarly, for $\hat{\mu}^{1}_{B}~\hat{\mu}^{3}_{C}$, we get,
\begin{equation*}
    \chi_{13}^{BC} = P^{C}_{q}/3.
\end{equation*}

Finally, to obtain the baryon contribution to the pressure, we take the linear combination,
\begin{equation}
    3\chi_{22}^{BC} - \chi_{13}^{BC} = 3 \left( \frac{P^{C}_{q}}{9} + P_{B}^{C} \right) - \left( \frac{P^{C}_{q}}{3} + P_{B}^{C} \right)
\end{equation}
which simplifies to,
\begin{equation}
    P_{B}^{C} = \frac{3}{2}\chi_{22}^{BC} - \frac{1}{2}\chi_{13}^{BC}.
\end{equation}
This final relation allows the partial pressure of baryons to be expressed entirely in terms of the susceptibilities $\chi_{22}^{BC}$ and $\chi_{13}^{BC}$, separating the baryon and quark contributions. Similarly, comparing the coefficient of the appropriate monomials, we can derive Eq.~\ref{equation_partial_pressure}-\ref{equation_partial_charged}.

\twocolumngrid

\end{document}